# Multimode Nonlinear Dynamics in Anomalous Dispersion Spatiotemporal Mode-locked Lasers

Yuhang Wu,[1,*] Demetrios N. Christodoulides[2], and Frank W. Wise[1]

[1]*School of Applied and Engineering Physics, Cornell University, Ithaca, New York 14853, USA*
[2]*CREOL/College of Optics and Photonics, University of Central Florida, Orlando, FL, USA.*
*\*Corresponding author: yw974@cornell.edu*

**Spatiotemporal mode-locking in a laser with anomalous dispersion is investigated. Mode-locked states with varying modal content can be observed, but we find it difficult to observe highly multimode states. We describe the properties of these mode-locked states and compare them to the results of numerical simulations. Prospects for the generation of highly-multimode states and lasers based on multimode soliton formation are discussed.**

Mode-locked fiber lasers have earned a sound reputation as compact, flexible, alignment-free, and power-scalable sources of femtosecond pulses. Mode-locked lasers with single-mode fiber (SMF) has been studied extensively in the past. People have utilized the linear and nonlinear dynamics to build novel saturable absorbers as well as to shape and manipulate the pulse [1]. Comparing to the pronounced control of single transverse mode fields, the attention paid to higher-order transverse modes is limited.

In recent years, there is an increased interest in nonlinear wave propagation in multimode fiber (MMF) [2-7]. With the two more spatial degrees of freedom compared to SMF, complicated space-time wave packets and their evolutions can be investigated. Many intriguing nonlinear multimode phenomena have been observed in this system [2-7]. These explorations are not only of scientific interest, but also may pave the way to generation of controllable spatiotemporal wave-packets in ways that would allow tailoring the group velocity and diffraction properties of the light field [8, 9].

Spatiotemporal mode-locking (STML) in multimode fiber lasers has attracted attention for the abundant new laser physics that is accessible as well as potential applications [10-16]. Wright et al. showed that high-order transverse modes can be self-organized into a mode-locked pulse under the principles of normal-dispersion mode-locking [10]. A follow-up theoretical study accounts for the main features of STML lasers to date and offers insight into the pulse-formation in the cavity [11]. Early follow-up work includes the observation of multimode soliton bound states [12], demonstration of self-similar temporal evolution in a STML fiber laser [13] and beam cleaning inside the laser cavity to produce a high-quality output beam [14]. STML has been realized with several different saturable absorbers [15, 16]. STML in a cavity with a segment of fiber with large modal dispersion has also been demonstrated [17]. All of these results were based on lasers with normal group-velocity dispersion.

Reports of STML in anomalous-dispersion lasers have just begun to appear. STML with anomalous dispersion can be regarded as the multimode analog of important and well-established soliton lasers based on SMF, but with richer nonlinear dynamics owing to linear and nonlinear intermodal interactions. Moreover, such lasers should provide another platform to investigate multimode solitons, which are not completely understood. The first anomalous-dispersion STML laser [18] generated picojoule-energy and 16-ps pulses by saturable-absorber mode-locking; the effects of dispersion and fiber nonlinearity on the pulse-shaping would be small. Zhang et. al. have observed mode-locking in multimode anomalous dispersion laser fiber cavities with a nonlinear multimodal interference saturable absorber [19, 20]. The output spectrum exhibited features expected for a (single-mode) soliton laser, and nearly transform-limited pulses between 300 fs and 2 ps were generated. These features suggest that soliton-like pulse-shaping may be playing a role in the lasers, but more work is needed to reach a conclusion.

Here we report a study of spatiotemporal mode-locking in lasers with anomalous dispersion multimode fiber. Observed mode-locked states are dominated by fundamental mode content. Numerical simulations qualitatively agree with experimental trends and reveal the intracavity dynamics. The results clarify the challenges that must be overcome to demonstrate mode-locking based on multimode soliton formation.

This experimental setup (Fig. 1) is designed to highlight multimode aspects of pulse propagation and facilitate comparison of experimental results with simulations. Graded-index (GRIN) MMF has small modal dispersion compared to step-index MMF, which reduces temporal walk-off and enhances nonlinear intermodal interactions. Graded-index gain fiber is not widely available, so we designed a cavity with segments of single-mode gain fiber and multimode graded-index passive fiber. We employ a unidirectional ring cavity. The single-mode section includes 1 m of passive fiber (HI1060) and 1 m of erbium-doped gain fiber (Thorlabs 8/125). The gain fiber is pumped by a single-mode pump diode. The GRIN fiber has 50 μm core diameter and supports 55 spatial modes in each of two polarizations, and is 2 m long. The gain fiber is spliced to the GRIN fiber with their cores offset to excite higher-order modes of the GRIN fiber. Multiple arcs are applied to reduce reflection from the splice. All of the fibers have anomalous dispersion at 1550 nm, and the total cavity dispersion is −0.09 ps$^2$. A spatiotemporal saturable absorber based on nonlinear polarization rotation is realized with the wave plates and polarizing beam splitter, which also serves as the output coupler. A birefringent plate is used as a spectral filter (21-nm bandwidth), which helps suppress noise bursts in single-mode soliton lasers [21]. The beam profile in the multimode fiber is characterized by imaging the output face of the fiber on a camera. To characterize the pulses at different positions on the beam, a pinhole that isolates the beam at the image plane is used. The pulse train and radio-frequency (RF) spectrum are recorded at each position, and the temporal profile at each position is characterized with an autocorrelator. Modal content is estimated roughly from the combination of the $M^2$ parameter and an approximate mode decomposition. The beam is sent into a multimode fiber and a spectrometer to record the spatially-integrated spectrum.

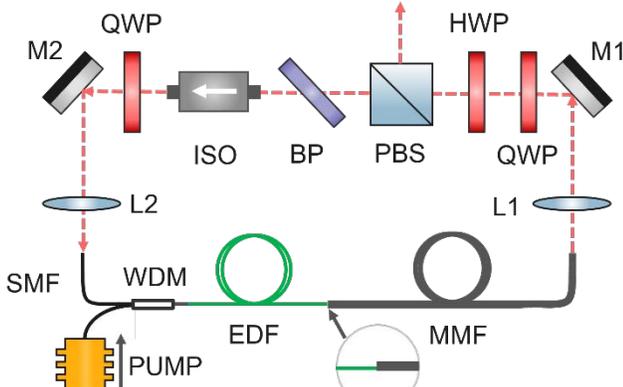

Figure 1. Schematic of cavity of the anomalous dispersion STML laser. SMF, single-mode fiber; WDM, wavelength division multiplexer; EDF, erbium-doped single-mode fiber; MMF, multimode fiber; L1, L2, collimating lens; M1, M2, mirrors; QWP, quarter-wave plate; HWP, half-wave plate; PBS, polarization beam splitter; BP, birefringence plate; ISO, isolator.

Mode-locked states are found by first optimizing the continuous-wave power (which is sensitive to the coupling between multimode and single-mode fibers), and then monitoring the pulsed output as the wave plates and coupling between fibers are adjusted. The pump power is increased, and the process is repeated. In typical mode-locked states, the laser generates ~300-fs pulses with ~0.8-nJ pulse energy. Further increases of the pump power lead to double-pulsing states. Once mode-locked states are found, they are typically stable for hours unattended.

The study of STML states with different modal populations should help to understand STML. The output beam profile and optical spectrum for splice offsets of 0, 6, and 10 μm are shown in Fig. 2. (A detailed characterization is below.) The decompositions of the beam profiles into the lowest 6 modes of the fiber (LP$_{01}$, LP$_{11a}$, LP$_{11b}$, LP$_{02}$, LP$_{21a}$, LP$_{21b}$) were obtained by assuming a flat phase front in the near field. We find that this method, while crude, can be used to estimate the mode contents of beams with a small number of features ("speckles") with reasonable accuracy. With increasing splice offset, the fundamental mode becomes less dominant. It is difficult to find stable mode-locked states with splice offset larger than 10 μm or high multimode content. It is worth mentioning that the observation of mode-locking with modest higher-order mode content is significant, considering that a tiny amount of energy in higher-order modes is known to destabilize mode-locking in lasers designed to work in a single transverse mode [22].

A signature of soliton lasers is the appearance of spectral sidebands. The laser with zero splice offset exhibits sidebands that qualitatively appear to be Kelly sidebands, and their positions agree with analytical theory [23, 24] as well as numerical simulations of a laser with single-mode fiber. The experimental cavity dispersion is used in the calculations and the simulations. Furthermore, the time-bandwidth product (0.33) is very close to the theoretical value for a soliton (0.315). Thus, we conclude that the laser with zero splice offset is a soliton laser operating in the fundamental mode of the MMF. This analysis provides the chromatic dispersion for the lasers made with offset splices. With increasing splice offset, we observe that the sharp Kelly sidebands disappear and are then replaced by other structures (marked in Fig. 2(f)). Known sources of resonant sidebands including Kelly sidebands [23, 24] and four-wave-mixing based sideband dips [25] do not account for these structures, which are reproducible in experiments. We do not understand the observed spectral structures, which presumably arise from intermodal energy-transfer processes.

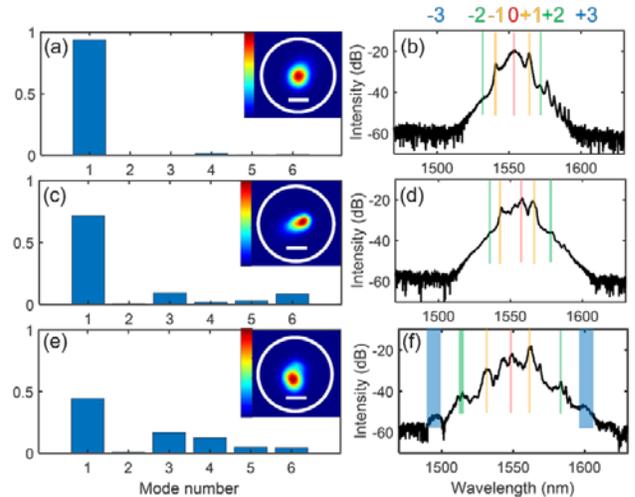

Figure 2. Comparing STML mode and spectrum information with different splicing shift. (a) and (b) are for 0 μm splicing shift. (c) and (d) are for 6 μm splicing shift. (e) and (f) are for 10 μm splicing shift. (a), (c), and (e) are the mode-decomposition results assuming a flat phase front (first 6 modes are shown). The inner patterns are near-field intensity patterns. (b), (d), and (f) are the integrated spectrum with sidebands marked with lines. The widths of the lines indicate the uncertainty in the peak positions.

The characterization of the mode-locked state with 10-μm splice offset, which yields the greatest multimode content, is shown in Fig. 3. The beam profile (Fig. 3(a)) is a single lobe with an elliptical shape and some additional structure. For vertical and horizontal directions, the $M^2$ values are 2.83 and 2.02 respectively (details are in supplementary material). These values are consistent with energy distributed among the lowest 6 modes. Autocorrelation measurements from two different points on the beam (Figs. 3(b,c)) show similar pulse shapes, with about 25% difference in pulse duration (310 fs and 250 fs); this state is less complex spatiotemporally than states observed in normal-dispersion STML [10, 17]. Strong evidence of STML comes from the RF spectra from two positions on the beam, which exhibit reasonably-high contrast and overlap completely (Fig. 3(d)). The dispersion length of the propagating pulse is comparable to the length of fiber in the laser, so the anomalous dispersion clearly plays a role in the pulse evolution.

Numerical simulations were performed to gain insight into the mechanism of anomalous dispersion STML. The pulse propagation is simulated using the generalized multimode nonlinear Schrodinger equation [26] with a massively parallel algorithm [27]. Parameters in the simulation are the experimental values. To efficiently model the cavity, the first 6 modes are included in the MMF part of the simulation. A few simulations performed with 10 modes yield qualitatively similar results. An effective saturable absorber is modeled by an ideal transfer function which is applied directly to the full field. The output coupling is taken to be 50%. Converged (i.e., stable) solutions are found for a reasonable range of parameters. A representative solution for 10-μm splice offset is shown in Fig. 4. The output beam (Fig. 4(a)) has a main lobe that deviates from the center of the fiber, with a small amount of energy in two other small lobes. (Other example simulations with different splice offsets are in the supplementary material.) The modes overlap temporally and the pulse duration in each mode is around 400 fs (Fig. 4(b)). In the spectral domain (Fig. 4(c)), all the modes have similar shapes, with little variation of the peak wavelength and no indication of structures that might be resonant sidebands. The temporal profiles and the lack of spatiotemporal complexity are qualitatively consistent with experimental results. However, the simulated spatial profile has more energy in higher-order modes than the experimental beam.

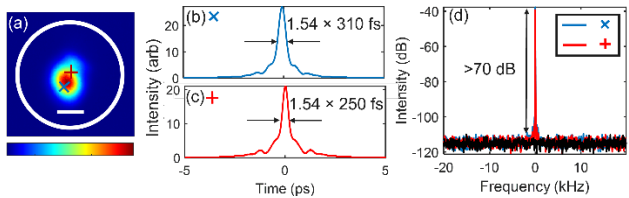

Figure 3. Experimental characterization of the 10 um-offset spliced anomalous STML laser. (a) Near-field beam intensity profile. The white circle is the core boundary. The white bar indicates the mode field diameter of the fundamental mode. (b), (c) autocorrelation traces at two points on the beam (marked in the (a)). (d) RF spectra from two points on the beam. The fundamental repetition rate is 37 MHz. The black curve is the noise floor.

Our interpretation of the simulation results is that a pulse forms as a result of the processes that underlie multimode soliton formation in passive fiber [4, 28], but does not reach a steady-state soliton solution (which would be perturbed by the gain and loss in the cavity). That would require all modes to propagate at the same velocity. We see that the modes separate temporally less than they would in linear propagation, but still continue to separate, in the GRIN fiber (Fig. 4(e)). The modes shift spectrally (modes 1-3 to the red, and modes 4 and 5 to the blue, Fig. 4(f) as a result of nonlinear coupling, to compensate the differences in modal velocities – the process that underlies MM soliton formation. The spatially integrated spectrum (Fig. 4(d)) shows that there are no apparent sidebands when many modes have significant energy content. This suggests that the experimentally-observed spectral structures are a consequence of the fact that the fundamental mode dominates the field.

The few-hundred femtosecond pulse duration in the anomalous-dispersion lasers is much shorter than the few-picosecond duration of chirped dissipative solitons that form with normal dispersion [10]. As a result, modal dispersion is a bigger challenge in anomalous-dispersion lasers than in normal-dispersion lasers. A spatial filter in the cavity can play a critical role in overcoming modal dispersion in normal-dispersion lasers [11]. In the lasers described here, the core of the SMF is a spatial filter and it does play a role in compensating the modal dispersion (Fig. 4(e)). A highly-multimode beam will suffer large loss on coupling into the SMF, and this limits the range of modes that can be synchronized temporally. We believe that this explains the difficulty in observing highly-multimode states experimentally. The use of gain fiber that supports a few modes should allow mode-locked states with greater content in higher-order modes to be found. It is interesting to note that Guo et al. report highly-multimode output from a laser with anomalous dispersion [20]. The speckled output beam occurs with optical spectra that have clear Kelly sidebands at the spectral positions predicted by the dispersion of the fundamental mode (Supplementary Information). Based on the appearance of the Kelly sidebands, our tentative understanding of the results of Ref. 20 is that the pulse propagation is dominated by soliton formation in the SMF that makes up much of the laser. However, we don't completely understand the generation of the highly-speckled beam in the multimode output coupler nor the conversion back to the SMF. A detailed understanding of the intracavity pulse evolution in this laser will be interesting.

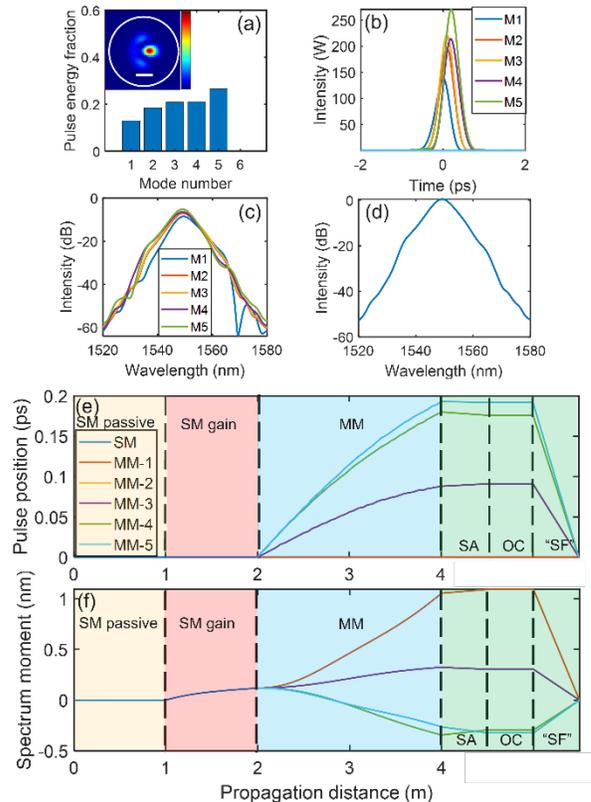

Figure 4. Simulation results for a cavity with 1 m passive SMF, 1 m gain SMF, and 2 m MMF. (a) Beam profile and modal energy fraction at the output of the multimode fiber. (b) Mode-resolved temporal profiles. (c) Mode-resolved spectra. (d) the spatially-integrated spectrum. (e) Pulse position relative to the fundamental pulse. SA: saturable absorber; OC: output coupler; SF: spatial filter effect caused by the single-mode fiber end. SM: mode in single-mode fiber; 'MM-1', 'MM-2', 'MM-3', 'MM-4', 'MM-5' are the first five modes in MMF. (f) First moment of the spectra relative to the spectrum in the fundamental mode.

The possibility of generating multimode solitons in a fiber laser was explored in simulations. The simulated cavity is designed to reduce the nonlinear pulse shaping in the SMF, increase the pulse energy, and include a long fiber aiming at forming stable MM solitons. It includes 0.5 m of single-mode gain and either 6 m or 8 m of passive GRIN MMF. The fibers are spliced with a small 2 μm offset. Starting from a 3 ps and 5 pJ gaussian pulse, the simulation for the cavity with 6 m of GRIN fiber converges to a stable solution. However, stable MM soliton propagation is still not observed. With GRIN fibers between 6 and 8 m long (the longest considered), the solution either does not converge or the converged solution is not a MM soliton. Fig. 5 shows the mode-resolved intensity profiles at the end of the GRIN fiber after different numbers of cavity round-trips. MM solitons typically form as energy corresponding to the unshifted spectral components disperses temporally [28]. Figs. 5(a-d) show the mode-resolved intensity profiles at the end of 6 m GRIN fiber. With 8 m of GRIN fiber (Fig. 5(e-h)), large separation of pulses causes multi-peak feedback into the SMF and disrupts the convergence process. As previous work emphasized, the spatial filter is important in resetting the temporal walk-off between modes [11]. In the lasers described here, this resetting ability is limited. Careful mode selection for the coupling between MMF and SMF may allow generation of multimode solitons inside the cavity. A laser based entirely on MMF would be interesting to study, but will require GRIN gain fiber. The Mamyshev mechanism [29] may offer another approach, based on controlling the temporal domain through spectral filtering.

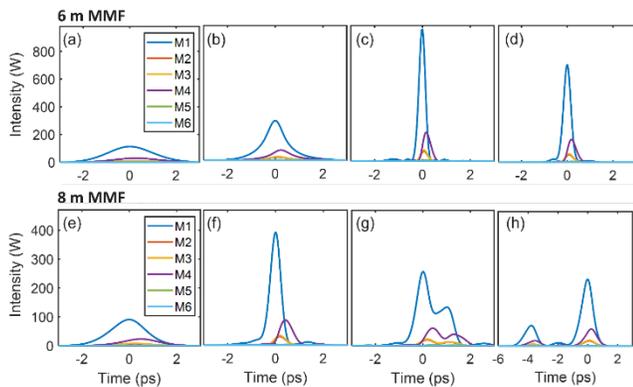

Figure 5. Simulation results for 0.5 m gain SMF and 6 m or 8 m MMF. (a-d) With 6 m MMF, the mode resolved temporal profile at the output of the MMF. The four panels correspond to 3, 5, 7, and 30 round trips. (e-h) With 8 m MMF, the mode resolved temporal profile at the output of MMF after 3, 5, 7, and 30 round trips.

In conclusion, we have numerically and experimentally studied an STML laser with anomalous dispersion. Mode-locked states with single-lobe intensity profiles and little spatiotemporal complexity are observed. Simulations show that the pulse-shaping is (multimode) soliton-like, but steady soliton propagation is not reached. The simulations indicate that mode-locked states with greater multimode content should be possible. Future directions will include the generation of multimode solitons inside the cavity and scaling the stable pulse energy.

**Acknowledgments**. The authors thank Henry Haig for valuable discussions.

**Disclosures**. The authors declare no conflicts of interest.

**Data Availability**. Data underlying the results presented in this paper are not publicly available at this time but may be obtained from the authors upon reasonable request.

**Supplemental document**. See Supplement 1 for supporting content.

# Multimode Nonlinear Dynamics in Anomalous Dispersion Spatiotemporal Mode-locked Lasers: Supplemental Document

## 1. Measurement results for lasers with zero and 6-µm splice offset

In the paper, the full characterization of the laser with 10-µm splice offset between single mode fiber (SMF) and multimode fiber (MMF) is shown. Here, we show a full characterization for the lasers with zero and 6-µm splice offset. Fig. S1 shows the characterization of output beam with 0-µm offset splice. An output pulse energy of ~1 nJ is obtained with 512 mW pump power. The mode-locked state is self-starting. The multi-pulsing state is reached at the pump power of 525 mW. Fig. S1(a), (b), and (c) are the near-field intensity profile, mode-decomposition assuming a planar phase front, and spatially-integrated spectrum respectively. Fig. S1(d) is the pulse measurement at the center of the beam. Fig. S1(e) is the measured pulse train. In this case, little energy populates modes other than the fundamental mode, so pulse measurements at different points lead to a similar pulse width. Fig. S2 shows the characterization of output beam with 6-µm splice offset. The pulse energy of ~0.9 nJ is obtained with pump power about 512 mW. Similar multi-pulsing threshold as for the zero-offset case is observed. Fig. S2(a), (b), and (c) are the near-field intensity profile, mode-decomposition assuming a planar phase front, and spatially-integrated spectrum respectively. Fig. S2(d) and (e) are the pulse measurements at the two spatial points of the beam. Fig. S2(f) is the measured pulse train.

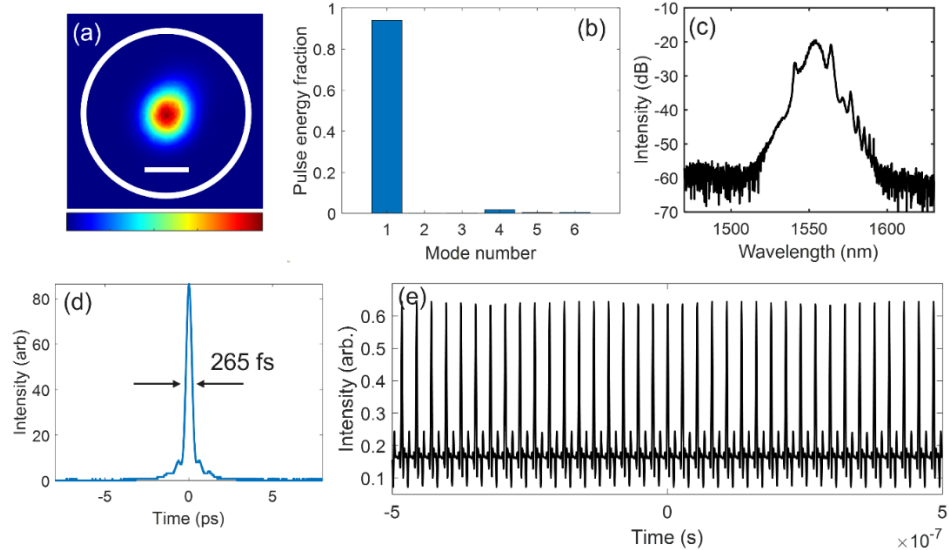

Figure S1: Mode-locked state with zero splice offset. (a) near-field intensity profile of output beam. (b) mode-decomposition result assuming a flat phase front. (c) spatially-integrated spectrum. (d) pulse autocorrelation measurement. (e) pulse train measured by a photodiode.

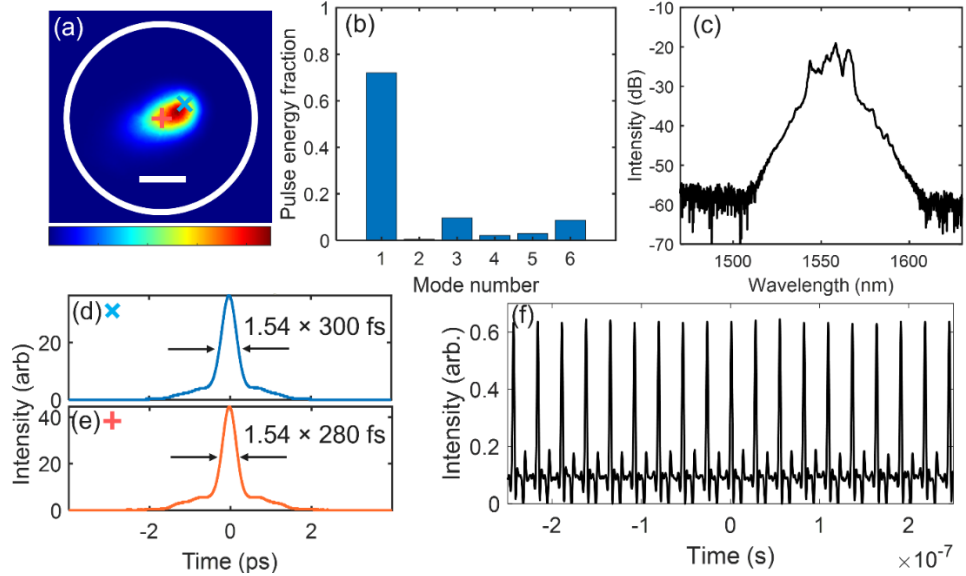

Figure S2: Mode-locked state with 6-µm splice offset. (a) near-field intensity profile of output beam. (b) mode-decomposition result assuming a flat phase front. (c) the spatially-integrated spectrum. (d) and (e) are the pulse autocorrelation measurements at two different positions of the beam. (f) pulse train measured by a photodiode.

## 2. M² measurement for zero and 10-µm splice offsets

To verify the mode contents of the output beam, M² measurements were carried out. The output beam from the laser is sent to a moving stage with a camera on it. Beam profiles at different positions are recorded. The D4σ parameters for the two transverse x and y directions are calculated from the measured beam profiles with Equation (S1). $\bar{x}$ and $\bar{y}$ are the averages in the x and y directions. The data are fit to Equation (S2) to determine M².

$$D_{4\sigma i} = 4\sqrt{\frac{\iint I(x,y)(i-\bar{i})^2 dxdy}{\iint I(x,y)dxdy}} \quad (i = x, y) \tag{S1}$$

$$(D_{4\sigma i}/2)^2 = W_0^2 + M^4\left(\frac{\lambda}{\pi W_0}\right)^2 (z - z_0)^2 \tag{S2}$$

Here, $W_0$ is the beam waist at $z_0$ and $\lambda$ is the wavelength. With zero splice offset $M_x^2 = 1.66$ and $M_y^2 = 1.88$ (Fig. S3(a)). The splice will inject a small amount of energy into the radially-symmetric LP$_{02}$ mode. Moreover, LP$_{01}$ itself is not a perfect Gaussian. Thus, we expect an M² value a bit larger than 1. With the 10-µm splice offset $M_x^2 = 2.02$ and $M_y^2 = 2.83$ (Fig. S3(b)).

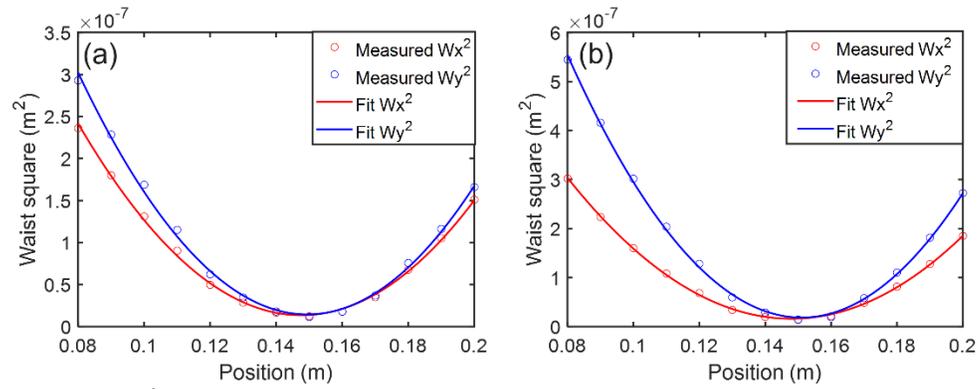

Figure S3. $M^2$ measurements for the output beam. (a) measurement with zero splice offset. $W_x$ is half of D4σ in x direction; $W_y$ is half of D4σ in y direction. (b) measurement with 10-μm splice offset.

### 3. Simulation results for different splice offsets

Simulation results for different values of splice offset are shown in Fig. S4. All of the beam profiles show one off-center lobe dominating. These beam patterns have the same features as what is observed in experiments (Fig. S1(a) and Fig. S2(a)). With the increase of splice offset, there is an increase in the energy populating higher-order modes (Fig. S4(b), (f), (j), (n), and (r)). Near-field intensity profiles are shown in Fig. S4(a), (e), (i), (m), and (q). The pulse width for different modes in different cases varies in the range of 350 fs to 700 fs (Fig. S4(c), (g), (k), (o), and (s)). Mode-resolved spectra are shown in Fig. S4(d), (h), (l), (p), and (t).

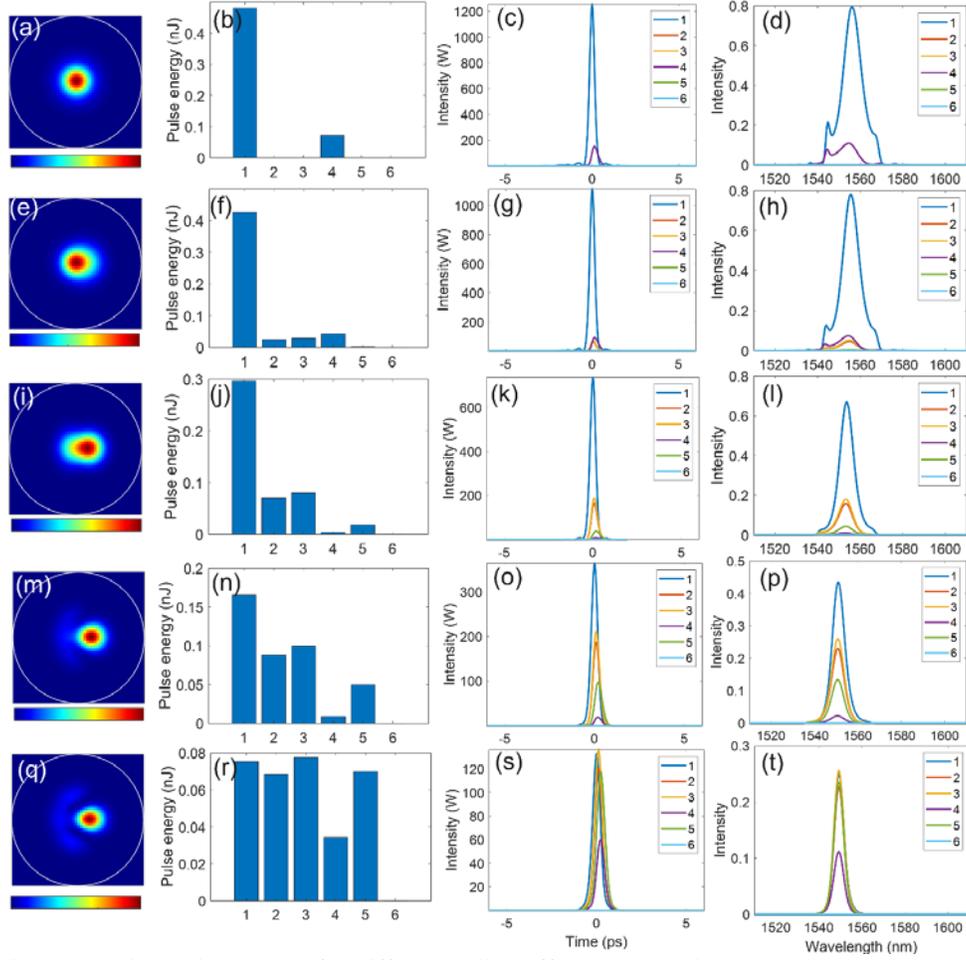

Figure S4. Simulation results for different splice offset. (a), (e), (i), (m), and (n) are the near-field intensity profiles. (b), (f), (j), (n), and (r) are the mode decompositions. (c), (g), (k), (o), and (s) are the mode-resolved temporal profiles. (d), (h), (l), (p), and (t) are the mode-resolved spectral profiles. (a), (b), (c), and (d) correspond to zero splice offset. (e), (f), (g), and (h) correspond to the 2-µm splice offset. (i), (j), (k), and (l) correspond to the 4-µm splice offset. (m), (n), (o), and (p) correspond to the 6-µm splice offset. (q), (r), (s), and (t) correspond to the 8-µm splice offset.

## 4. Calculation of the Kelly sideband positions in prior work

Guo et al. report observations of spatiotemporal mode-locking in a laser with anomalous dispersion [1]. The laser produces a highly-speckled output beam. In addition, the optical spectrum exhibits what appear to be Kelly sidebands. The frequencies of Kelly sidebands can be calculated with pulse width $t_p$, soliton period $Z_0$, and the amplifier period $Z_a$ (Equation (S3)) [2].

$$\delta v_n = \pm \frac{1}{2\pi t_p}\sqrt{-1 + 8nZ_0/Z_a} \qquad (S3)$$

Parameters of the laser in Ref. 1 are provided. The pulse width (FWHM), total dispersion of the cavity (using dispersion of fundamental mode of the MMF) and the total cavity length are

750 fs, -0.18 ps$^2$, and 13.3 m respectively. Based on those values, the first two sideband positions are calculated to be 10 nm and 15 nm away from the center wavelength. These values are very close to the sideband positions shown in Fig. 3 of Ref. 1.